\documentclass[conference]{IEEEtran}

\usepackage{color,graphicx}
\usepackage{subfigure}
\usepackage{amsmath}
\usepackage{amsfonts}
\usepackage{bm} 
\usepackage{epstopdf}

\usepackage{multirow}
\usepackage{rotating}

\begin{document}

\title{Massive MIMO and Waveform Design for\\5th Generation Wireless Communication Systems}

\author{\IEEEauthorblockN{Arman Farhang\IEEEauthorrefmark{1},
Nicola Marchetti\IEEEauthorrefmark{1},
Fabr\'{i}cio Figueiredo\IEEEauthorrefmark{2} and
Jo\~{a}o Paulo Miranda\IEEEauthorrefmark{2}}
\IEEEauthorblockA{\IEEEauthorrefmark{1}CTVR - The Telecommunications Research Centre, Trinity College, Ireland. Email: [farhanga, marchetn]@tcd.ie}
\IEEEauthorblockA{\IEEEauthorrefmark{2}CPqD - Research and Development Center on Telecommunication, Brazil.
Email: [fabricio, jmiranda]@cpqd.com.br}}

\maketitle
\begin{abstract}
This article reviews existing related work and identifies the main challenges in the key 5G area at the intersection of waveform design and large-scale multiple antenna systems, also known as Massive MIMO.
The property of self-equalization is introduced for Filter Bank Multicarrier (FBMC)-based Massive MIMO, which can reduce the number of subcarriers required by the system.
It is also shown that the blind channel tracking property of FBMC can be used to address pilot contamination -- one of the main limiting factors of Massive MIMO systems.
Our findings shed light into and motivate for an entirely new research line towards a better understanding of waveform design with emphasis on FBMC-based Massive MIMO networks.
\end{abstract}
%\vspace{-1.5mm}

\section{Introduction}
\label{sec:intro}
%\vspace{-1mm}
Ongoing societal developments are changing the way we use communication systems.
On-demand data is increasingly being delivered over mobile and wireless communication systems.
These developments will lead to a big rise of traffic volume.
Applications, such as broadband telephony and Machine-Type Communications (MTC), have diverse needs that fifth generation (5G) systems will have to support, $e.g.$ stringent latency and reliability requirements, and a wide range of data rates \cite{Metis2012}.
One of the main challenges is to meet these goals while at the same time addressing the growing cost pressure.
A popular view is that the required increase in data rate will be achieved through \textit{combined gains} \cite{Andrews2014} in extreme network densification (to improve area spectral efficiency), increased bandwidth (by exploiting mmWaves and making better use of unlicensed spectrum), and increased spectral efficiency (through advances in Multiple Input Multiple Output (MIMO) techniques).

In the quest for bandwidth, particular challenges that need to be addressed in the context of 5G are fragmented spectrum and spectrum agility.
It is unlikely that these challenges can be satisfied using Orthogonal Frequency Division Multiplexing (OFDM), and new waveforms are required.
Some researchers have also started to question the working assumption of strict synchronism and orthogonality in cellular networks, as a way to relax strict time domain requirements in case of sporadic traffic generating devices ($e.g.$ MTC devices) or applications requiring ultra-low latency, such as the Tactile Internet \cite{Wunder2014}.

Waveform design is quite a hot topic now, as it sheds light into how candidate waveforms perform in cellular environments, and discusses how they fare regarding specific aspects of 5G systems.
In \cite{Banelli2014}, for instance, OFDM is compared to Filter Bank Multi-Carrier (FBMC), Time Frequency-packed Signaling (TFS), and Single-Carrier Modulation (SCM):
OFDM is preferred in terms of ease of hardware implementation; SCM is the best candidate to reduce latency and peak-to-average power ratio (PAPR); FBMC is most robust against synchronisation errors; all waveforms but TFS can be used in mm-Wave bands, and all can be adopted in a \textit{Massive MIMO} setup, $i.e.$ systems using arrays with at least an order of magnitude more antennas than conventional (Multi-user MIMO) systems \cite{rusek:2013}.

A consequence of the powerful signal processing enabled by the large number of antennas is that most of the scheduling and PHY layer control issues in general are automatically resolved in Massive MIMO systems -- which of course is not the case for Multi-user MIMO systems with just a moderate number of antennas.
In recent past, Massive MIMO has gained significant momentum as potential candidate to increase capacity in multi-user networks.
In the limit, as the number $M$ of antennas at the Base Station (BS) tends to infinity, the system processing gain tends to infinity.
As a result, the effects of both noise and multi-user interference are removed \cite{Marzetta2010}.
Massive MIMO also enables significant latency reduction on the air interface, a welcome feature for delay-constrained applications \cite{larsson:2014}.

In general, it appears that any modulation technique, either single- or multi-carrier, can be used in combination with large antenna arrays.
According to \cite{Banelli2014}, it is reasonable to foresee that a similar behavior with respect to the vanishing of inter-user and inter-symbol interference (ISI) can be observed for any modulation format when the number of receiving antennas $M$ is sufficiently high.
However, since not all waveforms have equal advantages, the benefits of large antenna arrays can make a certain Massive MIMO-specific waveform combination more attractive than others.
FBMC offers lower out-of-band (OOB) emissions, and allows less expensive and more flexible carrier aggregation than OFDM, but traditionally had the problem of non easy applicability of MIMO to it \cite{farhang:2011}.
By scaling up the number of antennas, the combination of Massive MIMO and FBMC can benefit from the former's gains while still retaining the good properties of the latter \cite{Farhang2014}.

In \cite{larsson:2014} it is pointed out that, due to the law of large numbers, the channel hardens so that each subcarrier in a Massive MIMO system will have substantially the same channel gain.
Such a property has been also reported in \cite{Farhang2014} in the context of FBMC, where the authors name it \textit{self-equalization}, and can lead to a reduction in the number of subcarriers required by the system.
Another advantage of Massive MIMO is that, thanks to its many spatial degrees of freedom, the same frequency band can be reused for many users.
This plus the channel hardening render frequency domain scheduling no longer needed.
Issues related to Massive MIMO and waveform design for 5G are discussed in Section II and III, respectively.
Section IV presents some recent results of the authors on the combination of Massive MIMO and FBMC.
Section V concludes the paper. 
\begin{table*}[t]
%\vspace{-3ex}
  \caption{Summary of Challenges \& Solutions in Large-scale Multiple Antenna Systems for 5G.}
%  \vspace{-2ex}
  \label{tab:m_mimo}
  \begin{center}
  \begin{tabular}{|c|l|ll|l|@{}c@{}|} \hline
\textbf{Research Area} & \multicolumn{1}{|c|}{\textbf{Issue}} & \multicolumn{2}{c|}{\textbf{Candidate Solutions}} & \multicolumn{1}{|c|}{\textbf{Shortcomings and ``Side Effects''}} & \multicolumn{1}{|c|}{\textbf{Refs}} \\ \hline & & & & & \\ [-1em] \hline
& & & & $\bullet$ Diminishes bandwidth & \\
& Antenna coupling & \multicolumn{2}{l|}{Multiport impedance matching RF circuits} & $\bullet$ Introduces ohmic losses & \cite{janaswamy:2002}-\cite{artiga:2014} \\
Antenna & & & & $\bullet$ Not fully understood for large $M$ & \\ \cline{2-6}
Aspects & & & & $\bullet$ Increases coupling effects & \\
& Front-back ambiguity & \multicolumn{2}{l|}{Dense multidimensional implementations} & $\bullet$ Limited to indoor environments & \cite{janaswamy:2002},\cite{moustakas:2000} \\
& & & & $\bullet$ 3D arrays have restricted usefulness & \\ \hline

& \multirow{2}{*}{Channel modeling} & \multicolumn{2}{|l|}{$\bullet$ Realistic empirical models} & \multirow{2}{*}{Currently under development} & \hspace{1.5pt}\multirow{2}{*}{\cite{gao:2013},\cite{zheng:2014}}\hspace{1.5pt} \\
Propagation & & \multicolumn{2}{|l|}{$\bullet$ Sophisticated analytical models} & & \\ \cline{2-6}
& Cluster resolution & \multicolumn{2}{|l|}{No solution known to date} & Open research question & \cite{rusek:2013} \\ \hline

& & \multicolumn{2}{|l|}{BS sends pilots to terminals via FDD} & Limited by the channel coherence time & \cite{Marzetta2010},\cite{jiang:2014} \\ \cline{3-6}
& CSI acquisition & \multicolumn{2}{|l|}{\multirow{2}{*}{Terminals send pilots to BS via TDD}}& Channel reciprocity calibration & \cite{jose:2011}-\cite{rogalin:2014} \\ \cline{5-6}
& & & & Pilot contamination problem & \cite{ngo:2012}-\cite{Mueller:2014} \\ \cline{2-6}

& & \multirow{5}{*}{Linear precoding methods} & $\bullet$ ZF & $\bullet$ Computationally heavy for large $M$ & \multirow{2}{*}{\cite{rusek:2013},\cite{delamare:2013b}} \\
& \multirow{6}{*}{Precoding (in the DL)} & & $\bullet$ MMSE & $\bullet$ Higher average transmit power & \\ \cline{4-6}
& & & \multirow{2}{*}{$\bullet$ MF} & $\bullet$ Has an error floor as $M$ increases & \multirow{2}{*}{\cite{rusek:2013}} \\
& & & & $\bullet$ Higher $M$ required for a given SIR & \\ \cline{4-6}
& & & $\bullet$ BD & Cost-effective strategies are needed & \cite{delamare:2013} \\ \cline{3-6}
Transceiver & & \multirow{3}{*}{Nonlinear precoding methods} & $\bullet$ DPC & Extremely costly for practical deployment & \cite{caire:2003} \\  \cline{4-6}
Design & & & $\bullet$ THP & \multirow{2}{*}{Increased complexity is hard to justify} & \cite{windpassinger:2006} \\
& & & $\bullet$ VP & & \cite{peel:2005} \\ \cline{2-6}

& \multirow{7}{*}{Detection (in the UL)} & \multirow{2}{*}{Iterative linear filtering} & $\bullet$ MMSE-SIC & \multirow{2}{*}{Computationally heavy for large $M$} & \multirow{2}{*}{\cite{liang:2008}} \\
& & & $\bullet$ BI-GDFE & & \\ \cline{3-6}
& & \multirow{2}{*}{Random step search methods} & $\bullet$ TS & \multirow{2}{*}{More complex than MMSE-SIC} & \cite{zhao:2007} \\
& & & $\bullet$ LAS & & \cite{sun:2009} \\ \cline{3-6}
& & \multirow{3}{*}{Tree-based algorithms} & $\bullet$ SD & Complexity is exponential in $M$ & \cite{jalden:2005} \\ \cline{4-6}
& & & \multirow{2}{*}{$\bullet$ FCSD} & $\bullet$ 1000x more complex than TS & \multirow{2}{*}{\cite{barbero:2008}} \\
& & & & $\bullet$ Best suitable for the $M\approx K$ case & \\ \hline

\multirow{3}{*}{Hardware} & Phase noise & \multicolumn{2}{|l|}{Smart PHY transmission and receiver algorithms} & Efficacy yet to be demonstrated & \multirow{2}{*}{\cite{larsson:2014}} \\ \cline{2-5}
& Power consumption & \multicolumn{2}{|l|}{Parallel, dedicated baseband signal processing} & Open research question & \\ \cline{2-6}
& Proof-of-Concept (PoC) & \multicolumn{2}{|l|}{Experimental assessments, testbeds \& prototypes} & Only basic capabilities demonstrated & \cite{shepard:2012} \\ \hline
  \end{tabular}
  \end{center}
%\vspace{-4ex}
\end{table*}

%\vspace{-1mm}
\section{Large-scale Multiple Antenna Systems}
\label{sec:challengesMMIMO}
%\vspace{-1mm}
This section discusses issues regarded as most challenging in the Massive MIMO literature.
Table \ref{tab:m_mimo} lists such issues and their available solutions, each presented alongside with its side effects, $i.e.$ new issues brought about by their adoption.
%

%\vspace{-1.5mm}
\subsection{Mutual Antenna Coupling and Front-back Ambiguity}
%\vspace{-1mm}
One assumption often made when modeling antenna arrays is that the separation among antenna elements is large enough to keep mutual coupling at a negligible level.
This assumption is not entirely realistic, especially if a large number of elements is to be deployed as an array of constrained size and aperture.
Under such practical conditions, mutual coupling is known to substantially impact the achievable system capacity \cite{janaswamy:2002}.
Multiport impedance matching RF circuits can cancel out such coupling effects \cite{wallace:2004}, but they diminish output port bandwidth \cite{lau:2006} and increase ohmic losses \cite[Chapter~10]{sibille:2010}.

Two- or three-dimensional arrays have been reported able to avoid front-back ambiguity.
A side effect of dense implementations is that the larger the number of adjacent elements, the larger the increase of coupling effects \cite{rusek:2013}.
Another fundamental shortcoming specific to 3-D settings is the incapability of extracting additional information from the elements inside the array, $i.e.$ only elements on the array surface contribute to the information capacity \cite{moustakas:2000}.
The optimal densities above which performance deteriorates no matter how large is the number of elements are studied in \cite{artiga:2014} for indoor Massive MIMO BSs.

%\vspace{-1mm}
\subsection{RF Propagation and Channel Modeling}
%\vspace{-1mm}
Realistic performance assessments call for appropriate channel characterization and modeling.
The Massive MIMO channel behavior, including its correlation properties and the influence of different antenna arrangements, cannot be captured otherwise.
The interest raised by this issue has been (and still is) experiencing a fast-paced growth, and the community has already managed to contribute towards a better understanding on the matter.
In \cite{gao:2013}, channel measurements are carried out to identify and statistically model the propagation characteristics of interest.
These are then fed back into an existing channel model, extending its applicability to large-scale antenna arrays.

Performance assessments should ideally be conducted using a standardized or widely accepted channel model.
At the time of writing, no such a model seems to exist for Massive MIMO.
See, $e.g.$ \cite{zheng:2014}, for a discussion on modeling methods, channel categories, and their underlying properties.

%\vspace{-1.5mm}
\subsection{Channel State Information, Precoding \& Detection}
%\vspace{-1.5mm}
In conventional systems, the BS cannot harness beamforming gains until it has established a communication link with the terminals.
Firstly, the BS broadcasts pilots based on which the terminals estimate their corresponding channel responses.
These terminal estimates are then quantized and fed back to the BS.
Such Frequency-division Duplexing (FDD) finds limited application in Massive MIMO systems in that the amount of time-frequency resources needed for pilot transmission in the downlink (DL) scales as the number of antennas, and so does the number of channel responses that must be estimated on the part of each terminal.
In large arrays, pilot transmission time may well exceed the coherence time of the channel \cite{Marzetta2010}\cite{jiang:2014}.

An alternative for Massive MIMO systems is to let terminals send pilots to the BS via Time-division Duplexing (TDD).
TDD relies on channel reciprocity, where uplink (UL) channels serve as estimate of DL channels.
This leads to training requirements independent of $M$ \cite{jose:2011}, and eliminates the need for channel state information (CSI) feedback.
TDD's drawbacks are reciprocity calibration and pilot contamination: the former is a need raised by different transfer characteristics of DL/UL; the latter arises in multi-user scenarios where the use of non-orthogonal pilot sequences causes the intended user's channel estimate to get contaminated by a linear combination of other users' channels sharing that same pilot.
Reciprocity calibration and pilot decontamination are studied in \cite{kaltenberger:2010}-\cite{rogalin:2014} and \cite{ngo:2012}-\cite{Mueller:2014}, but optimal solutions are unknown to date.

Multi-user interference can be mitigated at the transmit side by modifying standard single-stream beamforming techniques to support multiple streams.
Precoding based on Zero-Forcing (ZF) or Minimum Mean Square Error (MMSE) is simple for a number of antennas up to moderate.
However, its reliance on channel inversions may take complexity and power burdens to a point hard to accommodate within very large arrays \cite{rusek:2013}\cite{delamare:2013b}.
Matched Filtering (MF), $i.e.$ Maximum Ratio Transmission (MRT) in the DL and Maximum Ratio Combining (MRC) in the UL, is known to be the simplest method \cite{Marzetta2010}.

Dirty Paper Coding (DPC) \cite{caire:2003}, Tomlinson-Harashima Precoding (THP) \cite{windpassinger:2006}, and Vector Perturbation (VP) \cite{peel:2005}, also have appealing features (DPC is theoretically optimal) but are either too costly for practical deployment or offer gains hard to justify in view of their increased computational complexity.
Recalling that the array size required to achieve a given Signal-to-Interference Ratio (SIR) using MF is at least two orders of magnitude larger than using ZF precoding \cite{rusek:2013}, further work on cost-effective solutions is needed.
This is illustrated in \cite{delamare:2013} for the case of Block Diagonalization (BD) algorithms.

When it comes to separation of data streams in conventional systems, Maximum Likelihood (ML) detection is the optimal solution but its complexity grows exponentially with the number of streams.
This is the reason why parameter estimation and detection are key problems in Massive MIMO systems.
Suboptimal ZF, MMSE, and MF offer lower costs (that do not depend on the modulation), but are not capable of achieving the full receive-diversity order of ML detection \cite{delamare:2013b}.

This performance-complexity tradeoff led to methods, such as MMSE with Successive Interference Cancellation (SIC), Block-iterative Generalized Decision Feedback Equalization (BI-GDFE) \cite{liang:2008}, Tabu Search (TS) \cite{zhao:2007}, Likelihood Ascent Search (LAS) \cite{sun:2009}, and Fixed Complexity Sphere Decoding (FCSD) \cite{jalden:2005}\cite{barbero:2008}.
Repeated matrix inversions in MMSE-SIC and BI-GDFE may be computationally heavy for large arrays, and matrix-inversion free methods such as TS and LAS are both outperformed by MMSE-SIC \cite{rusek:2013}.
This is an indication that more work is needed on this matter, perhaps towards turbo codes or Low-density Parity-check (LDPC) codes in iterative detection and decoding settings \cite{delamare:2013b}.

%\vspace{-1mm}
\subsection{Impairments due to Low-cost Hardware }
%\vspace{-1mm}
Large-scale multiple antenna arrays will most likely be built using low-cost components so as to ease the introduction and leverage the penetration of the Massive MIMO technology into the market.
This calls for solutions capable of circumventing hardware imperfections that manifest themselves as I/Q imbalance or phase noise.
The latter issue is of particular concern because low-cost power amplifiers often have relaxed linearity requirements, which in turn translate into the need for reduced PAPR on a per antenna element basis \cite{larsson:2014}.

Savings in radiated power result from using excess antennas to simultaneously send independent data to different users, but the total power consumption should also be taken into account.
In this context, an interesting research path is hardware architectures for baseband signal processing \cite{larsson:2014}.
Another path of interest is experimentation, as testbeds currently available only demonstrate basic capabilities, and do not take constrained BS real estates into consideration \cite{shepard:2012}.

\begin{table*}[t]
%\vspace{-3ex}
  \caption{Candidate Waveforms Regarded as Most Promising for 5G with CP-OFDM used as Benchmark (TBI = remains to be investigated).}
%  \vspace{-2ex}
  \label{tab:waveform}
  \begin{center}
  \begin{tabular}{|@{}l@{}|l|c|c|c|c|c|c|c|} \hline
\multicolumn{2}{|c|}{\textbf{Figure of Merit}} & \textbf{CP-OFDM} & \textbf{NC-OFDM} & \textbf{DFT-s-OFDM} & \textbf{BFDM} & \textbf{FBMC} & \textbf{GFDM} & \textbf{UFMC} \\ \hline & & & & & & & & \\ [-1em] \hline
\multirow{10}{*}{\vspace{-0.1cm}\begin{sideways}Performance\end{sideways}} & Peak-to-average Power Ratio & High & High & Reduced & High & High & Reduced & High \\ \cline{2-9}
& Spectral Efficiency & Low & Low & Low & High & High & High & High \\ \cline{2-9}
& Overhead & High & High & Variable & Low & Low & Variable & Low \\ \cline{2-9}
& Frequency Localization & Good & Good & Very good & Controllable & Excellent & Excellent & Excellent \\ \cline{2-9}
& OOB Emissions & High & Reduced & Reduced & Variable & Negligible & Reduced & Reduced \\ \cline{2-9}
& Sidelobe Attenuation [dB] & 13 & 20-50 & 40-60 & 13-60 & 60 & 35 & 40-60 \\ \cline{2-9}
& Bit Error Rate & Good & Good & Good & Good & Good & Very good & Very good \\ \cline{2-9}
& Throughput & Low & Low & Low & High & High & High & High \\ \cline{2-9}
& Time Offsets Resiliency & Poor & Good & Good & Good & Good & Good & Good \\ \cline{2-9}
& Frequency Offsets Resiliency & Poor & Good & Good & Good & Good & Good & Good \\ \hline
\multirow{4}{*}{\vspace{-0.4cm}\begin{sideways}Feasibility\end{sideways}} & Computational Complexity & Low & Low & Low & High & High & High & High \\ \cline{2-9}
& Implementation & Efficient & Efficient & Efficient & TBI & Efficient & Efficient & Efficient \\ \cline{2-9}
& Equalization & Simple & Simple & Simple & TBI & Involved & Simple & Involved \\ \cline{2-9}
& \multirow{2}{*}{Resource Allocation} & Dynamic and & Dynamic and & Dynamic and & \multirow{2}{*}{Possible} & \multirow{2}{*}{Configurable} & Configurable & \multirow{2}{*}{Configurable}\\
& & fine grained & fine grained & fine grained & & & and adaptable & \\ \hline
\multirow{4}{*}{\vspace{-0.42cm}\begin{sideways}Support for\end{sideways}} & Conventional MIMO & Yes & Yes & Yes & TBI & No & Yes & Yes \\ \cline{2-9}
& High-order Modulation & Yes & Yes & Yes & TBI & TBI & Yes & TBI \\ \cline{2-9}
& Short-burst Traffic (MTC) & No & Yes & Yes & Yes & No & Yes & Yes \\ \cline{2-9}
%& Asynchronous Signaling & No & No & No & Yes & Yes & Yes & Yes \\ \cline{2-9}
& Fragmented Spectrum & No & Yes & Yes & Yes & Yes & Yes & Yes \\ \cline{2-9}
& Low Latency (Tactile Internet) & No & No & No & No & No & Yes & No \\ \hline
\multicolumn{2}{|l|}{\hspace{2.5mm}References} & \cite{Andrews2014}-\cite{Banelli2014},\cite{farhang:2011} & \cite{loulou:2013} & \cite{berardinelli:2013},\cite{berardinelli:2014} & \cite{kasparick:2014} & \cite{Andrews2014},\cite{Wunder2014},\cite{farhang:2011},\cite{renfors:2014} & \cite{michailow:2014},\cite{datta:2014} & \cite{Andrews2014},\cite{Wunder2014}\cite{schaich:2014} \\ \hline
      \end{tabular}
  \end{center}
%\vspace{-4ex}
\end{table*}

%\vspace{-1mm}
\section{Waveform Design for 5G}
%\vspace{-1mm}
This section presents the state of the art about candidate waveforms for 5G.
Comparisons in Table \ref{tab:waveform} are based on the best possible performance of each waveform, with OFDM with cyclic prefix (CP-OFDM) used as benchmark.

%\vspace{-1mm}
\subsection{The Baseline OFDM and its Enhancements}
%\vspace{-1mm}
Despite the advantages that led to the near-universal adoption of CP-OFDM, it is not without its limitations \cite{farhang:2011}.
High PAPR in Massive MIMO is a concern, as it sets up a tradeoff between the amplifier's linearity and cost.
MmWave deployment may also prove hard due to the difficulty to develop efficient amplifiers \cite{Andrews2014}.
Spectral efficiency \cite{Wunder2014} can be improved by means of shorter CP lengths, and Frequency and Quadrature Amplitude Modulation (FQAM) to boost DL throughput for cell-edge users.
TFS and faster-than-Nyquist signaling have been claimed able to offer efficiency gains on the order of 25\% over conventional OFDM (see \cite{Andrews2014}\cite{Banelli2014} and references therein).
Other drawbacks of CP-OFDM are sensitivity to phase noise and asynchronous signaling, poor spectrum localization, large OOB emissions, and long round-trip time (RTT).

The amount of implementation experience and knowledge about tricky aspects of OFDM available today make it possible to modify it to create new schemes capable of circumventing most of its inherent limitations.
Improved sidelobe suppression for dynamic spectrum access and fragmented spectrum use is claimed achievable using noncontiguous waveforms, such as Cancellation Carriers (CC) or Edge Windowing (EW).
In \cite{loulou:2013}, CC-OFDM with a single cancellation carrier is shown better than both its variant with multiple (weighted) cancellation carriers and EW-OFDM, but suppression performance degrades as the subcarrier index runs away from the gap edge.

Another option consists of manipulating OFDM to mimic SCM \cite{Andrews2014}\cite{Banelli2014} to reduce PAPR and provide robustness against frequency offsets.
Discrete Fourier Transform spread OFDM (DFT-s-OFDM) enhances noise in faded channels, offers poor spectral containment, and allows neither frequency-selective scheduling nor link adaptation.
These limitations are overcome by employing zero-tail DFT-s-OFDM, exploiting receiver diversity, and appying the DFT spread at the physical resource blocks level \cite{berardinelli:2013}\cite{berardinelli:2014}.
Improved flexibility (dynamic overhead adaptation instead of CP hardcoding) and OOB emissions (smoother transitions between adjacent symbols) are additional advantages of zero-tail DFT-s-OFDM over CP-OFDM.

%\vspace{-2mm}
\subsection{Filter Bank Multicarrier}
%\vspace{-2mm}
FBMC introduced multicarrier techniques over two decades before the introduction of OFDM in wireless communications systems \cite{farhang:2011}.
While OFDM relies on the CP to prevent ISI and to convert the channel into a set of flat-gain subcarriers, FBMC exploits the fact that narrow and numerous subcarriers can be characterized by a flat gain.
The length and superior frequency localization of FBMC prototype filters allow the terminal to deal with high delay spreads and compensate frequency offsets without feedback to the BS \cite{Andrews2014}\cite{Wunder2014} (at the expense of increased complexity, latency, and equalization requirements).

Fast-convolution based highly tunable multirate filter banks are investigated in \cite{renfors:2014}.
Capable of implementing waveform processing for multiple single-carrier and/or multicarrier transmission channels with nonuniform bandwidths and subchannel spacings simultaneously, this method is a competitive option in terms of spectral containment and complexity.

%\vspace{-2mm}
\subsection{``Born-to-be-5G'' Waveforms}
%\vspace{-2mm}
In contrast to FBMC or OFDM, which apart from enhancements like CC-OFDM and DFT-s-OFDM were not originally designed bearing 5G requirements in mind, we have recently witnessed the outbreak of waveforms crafted for MTC and the Tactile Internet.
Biorthogonal Frequency Division Multiplexing (BFDM) waveforms, for instance, have been regarded as suitable to support sporadic data traffic and asynchronous signaling.
One appealing feature of BFDM is time and spectral localization balancing through iterative interference cancellation, which in turn allows to control degradations due to time and frequency offsets \cite{kasparick:2014}.

Rendered attractive for nonsynchronous burst transmissions by its block-based structure, Generalized Frequency Division Multiplexing (GFDM) was originally proposed as a nonorthogonal alternative to FBMC.
GFDM can be set to mimic OFDM, although its benefits are better experienced with SCM setups, $e.g.$ to transmit multiple symbols per subcarrier.
To the best of our knowledge, GFDM is the only 5G candidate waveform for which support for High-order Modulation (HOM) and Tactile Internet has been explicitly investigated.
We refer the reader to \cite{michailow:2014}\cite{datta:2014} for a recent analysis of characteristics, relevant features, performance, and implementation aspects of GFDM.

%\vspace{-1.5mm}
\subsection{Universal Filtered Multicarrier}
%\vspace{-1.5mm}
Another distinguishing aspect of OFDM and FBMC is that the former applies filtering on the whole band, while the latter works on a per subcarrier basis.
Universal Filtered Multicarrier (UFMC) has been advanced as a more general solution because its filtering is applied on the level of multiple subcarriers, $e.g.$ on a per resource block basis.
As compared with OFDM, UFMC offers better spectral efficiency and robustness against time and frequency offsets \cite{Andrews2014}\cite{Wunder2014}.
Some advantages of UFMC over FBMC are lower latency (due to its shorter filter lengths), reduced overhead, and improved support for MTC \cite{Wunder2014}\cite{schaich:2014} -- although both may require more involved multi-tap equalizers.

%\vspace{-1mm}
\section{FBMC-based Massive MIMO Networks}
%\vspace{-1mm}
Networks resulting from the combination of Massive MIMO and FBMC are of the utmost importance as in these systems spectrum not only can be reused by all the users, $i.e.$ advantage of Massive MIMO, but can also be used in an efficient manner, $i.e.$ due to the FBMC's low OOB emissions.
The application of FBMC to Massive MIMO was first considered in \cite{Farhang2014}, with an interesting finding of this work being the \textit{self-equalization} property of FBMC in Massive MIMO channels (in contrast with the limited applicability of FBMC to conventional MIMO channels).
As a result, FBMC can leverage various benefits that place it in a strong position as a candidate for 5G systems.

Linear combining of the received signals in different receive antennas at BS averages channel distortions between the users and BS antennas.
As $M$ increases, the channel distortions over each subcarrier are smoothed through linear combining, so a nearly equalized gain across each subcarrier band can be achieved.
Analytical SINR relationships are derived in \cite{Farhang2014} for MMSE and MF linear combiners under the assumption of a flat channel over each subcarrier band.
They are therefore used as benchmark to evaluate channel flatness in what follows.

\begin{figure}[t]
\centering
\includegraphics[scale=0.7]{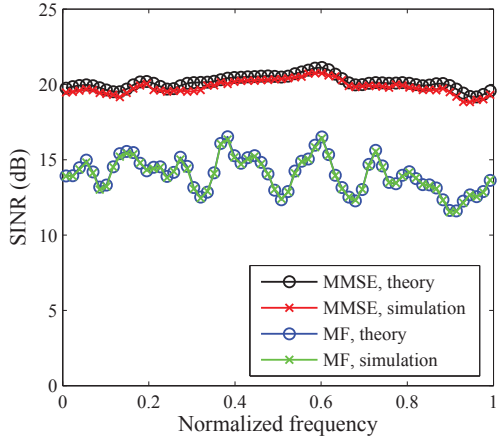}
%\vspace{-6 mm}
\caption{SINR comparison: MMSE vs. MF ($K=6$, $L=64$, $M=128$).}
\label{fig:SINR_MultiuserL32}
%\vspace{-3 mm}
\end{figure}

Figure \ref{fig:SINR_MultiuserL32} shows theoretical and simulation results of a multi-user scenario ($K=6$ users, $L=64$ subcarriers, and $M=128$ BS antennas).
The target output SINR of 20 dB may be calculated as SNR$_{\textup{in}}+10 \log_{10} M$, where SNR$_{\textup{in}}$ is the Signal-to-Noise Ratio (SNR) at each BS antenna, and $10\log_{10} M$ is the spreading gain due to $M$ BS antennas.
SINRs are evaluated over all subcarrier channels, and the number of points along the normalized frequency is equal to the number of subcarrier bands, $L$.
MMSE is superior to MF, and its SINR is about the same for all subcarriers, $i.e.$ has smaller variance across the subcarriers.
Our simulation results match very well to the theoretical ones, confirming the self-equalization property of linear combining in FBMC-based Massive MIMO systems.

This self-equalization property of FBMC relaxes the large $L$ requirement to obtain an approximately flat gain over each subcarrier band, so wider subcarriers can be used.
The use of a smaller $L$ in a given bandwidth: (1) reduces the latency caused by synthesis/analysis filter banks; (2) improves bandwidth efficiency due to the absence of the CP and to shorter preambles; (3) decreases computational complexity due to the smaller FFT and IFFT blocks needed for implementation; (4) provides robustness to frequency offsets; and (5) reduces PAPR.

As recently highlighted in \cite{Farhang2014_1}, Cosine Modulated Multitone (CMT), $viz.$ a particular FBMC form, has a blind equalization capability \cite{Farhang2003}, which can be used to decontaminate erroneous channel estimates in multicellular Massive MIMO networks caused by pilot contamination.
This approach, which is somewhat similar to the Godard blind equalization algorithm \cite{Godard1980}, can be easily extended from single antenna to Massive MIMO systems.
To alleviate the performance loss caused by corrupted channel estimates, the blind equalization technique adaptively corrects the linear combiner tap weights.
Starting with MF using noisy channel estimates, the SINR performance reaches that of MF with noise-free channel matrix within a small number of iterations, and keeps improving to get to that of MMSE with noise-free channel matrix.

\begin{figure}
\centering
%\vspace{-3.5 mm}
\includegraphics[scale=0.64]{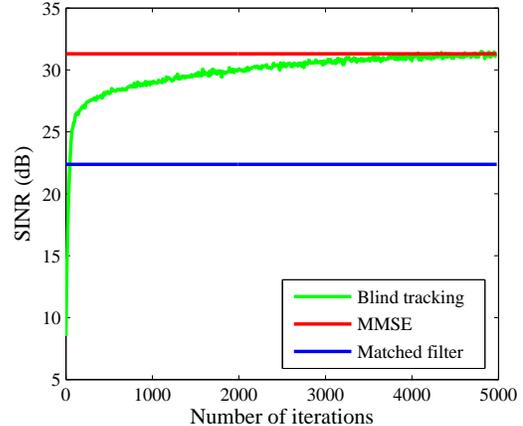}
%\vspace{-7 mm}
\caption{SINR comparison: Blind tracking technique vs. MF vs. MMSE ($L=256$, $M=128$).}
\label{fig:SINR}
%\vspace{-6 mm}
\end{figure}

This is shown in Figure~\ref{fig:SINR} for a Massive MIMO network in TDD mode with seven cells and one user per cell.
All users are assumed to use the same pilot sequences, $L=256$, and $M=128$.
In some situations where the length of the UL data packets is close to the channel coherence time, the estimated CSI in the beginning of the packet may get outdated, resulting in a performance loss (as the same CSI is used for precoding in the DL).
This problem can be alleviated by the utilization of blind channel tracking techniques as the one mentioned above, since these provide up-to-date CSI.
In other words, the blind channel tracking techniques sweep through all the symbols in the data packet and update the CSI.
The latest CSI taken from the last transmitted symbols can thus be obtained.

%\vspace{-3mm}
\section{Conclusion}
\label{sec:concl}
%\vspace{-2mm}
This paper has reviewed existing work and identified main challenges at the intersection of Massive MIMO and waveform design.
The property of self-equalization was introduced for FBMC-based Massive MIMO systems, which can help reduce the number of subcarriers required by the system.
It is also shown that the blind channel tracking property of FBMC can be used to address pilot contamination -- a major limiting factor of Massive MIMO systems.
Our findings shed light into and motivate for an entirely new research avenue towards a better understanding of waveform design for 5G with a particular emphasis on FBMC-based Massive MIMO networks. 

\section*{Acknowledgment}
We acknowledge support from the Science Foundation Ireland under grant No. 10/CE/i853 and the Irish Research Council New Foundations 2013 scheme under grant IB-COM.

%\vspace{-2mm}

\end{document}